\documentclass[12pt,a4paper]{article}
\begin{document}

\large

\date{}

\title{On solvability of the Cauchy problem for a second \\ order
parabolic equation
degenerating \\ into Schrodinger type }

\author{H.~I.~Ahmadov \footnote{E-mail: hikmatahmadov@yahoo.com} \\
Department of Mathematical Physics\\
Faculty of Applied Mathematics and Cybernetics\\
Baku State University, Z.Khalilov st.23, AZ-1148\\
Baku, Azerbaijan}

\maketitle

\begin{abstract}
The Cauchy problem is investigeted for the parabolic type in the
some finite part $\left[ t_0,t_1\right] \subset \left[ 0,\infty
\right)$ of the semi axis $t\in \left[ 0,\infty \right)$ and
degenarated to Schrodinger type in the remain part of the same
semi axes the second order parabolic equation.

The existence of the solution is proved under some conditions on
the data and the explicit integral representation is constructed
\end{abstract}

In the semi plane $\Pi =\left\{ (t,x)/t>0;\,-\infty <x<\infty
\right\}$ we consider the following Cauchy problem

\begin{equation}
p(t)\frac{\partial u}{\partial t}=\frac{\partial ^2u}{\partial
x^2}+ f(x,t),\qquad (t,x)\in \Pi ,
\end{equation}
\begin{equation}
\lim\limits_{t \to +0} u(t,x)= \varphi(x),\qquad -\infty <x<\infty
\end{equation}
where $p(t), f(t,x)$ and $\varphi(x)$ are the known functions,
$u=u(t,x)$ -are the desired complex valued functions.

 Relatively to the coeffisients and right hand sides of problem (1),
(2) it
is assumed the fulfilment of the following conditions:

1$^0$.\,\,\,$p(t)\in C\left[0,\infty \right),$

2$^0$.\,\,\,$p(t)\neq 0,\,\,\,\,\,$ at $\,t\in \left[0,\infty
\right),$

3$^0$.\,\,$Re\,p(t)\geq 0,\,\,\,\,$ at $\,\,\,t\in \left[ 0,\infty
\right)$ and $\,\,\,\,Re\,\,p(0)>0,$

4$^0$. There exists $p_0=const >0$ such that
$\int\limits_{0}^{t}Jmp^{-1}(\tau )\,\,d\tau \leq p_0,$

5$^0$. The function $\,\,\varphi (x)$ is continuous and bounded at
$x\in (-\infty ,\infty )$

6$^0$. {\it f(t,x)}\,\, is continions and bounded in the layer
$\Pi'(t_0,T)= \\
\hspace*{1.2cm} \left\{t_0\leq t\leq
T;\,\,\,-\infty <x<\infty \right\}$

7$^0$. {\it f(t,x)} satisfies in $\Pi'(t_0,T)$ the Holder
condition wish respect to $x$ i.e. \\
\hspace*{1.2cm} there exist the constants $B$ and $0< \alpha \leq
1$ such that $|f(t,x)- \hspace*{1.2cm} f(t,y)| \leq
B|x-y|^{\alpha}$ for any $(t,x),(t,y)\in \Pi'(t_0,T)$.

The formal solution of problems (1), (2) is constructed the help of
method of integral Fourier transform and is represented in the
form of

\begin{equation}
u(t,x)=\int\limits_{-\infty}^{\infty} Q(t,y-x)\varphi(y)\,dy +
\int\limits_{0}^{t} \int\limits_{-\infty}^{\infty}
Q_0(t-\tau,y-x)f(\tau,y)\,d\tau\,dy
\end{equation}

where

$$
Q(t,y-x)=\frac{e^{-\frac{(y-x)^2}{4\omega (t)}}}{2\sqrt{\pi
\omega(t)}} ,\,\,\,\,\,\,\,\,Q(t-\tau
,y-x)=\frac{e^{-\frac{(y-x)^2}{4\omega _0(t,\tau )} }}{2\sqrt{\pi
\omega _0(t,\tau )}}\,
$$

$$
\omega (t)=\int\limits_{0}^{t}p^{-1}(\eta )\,d\eta
,\,\,\,\,\,\,\omega _0(t,\tau )=\int\limits_{\tau}^{t}
p^{-1}(\eta)\,d{\eta} .
$$
Note that equation (1) in some part of the considered interval
$\,t\in \left[ 0,\infty \right)$, belongs to the parabolic type,
in the other parts of interval is degenerated in Schrodinger type
[1].

Let $\left[ t_0,t_1\right] \subset \left[ 0,\infty \right)$ be a
segment, where the condition

\begin{equation}
Re\,\,p(\tau )>0,\qquad \tau \in \left[ t_0,t_1\right]
\end{equation}

is satisfied.

At fulfilment of conditions $1^0-4^0$, some estimetes for the elements
of
integral (3) which provide uniform convergence of this integral, are
obtained.

The sollowing one is proved.

Lemma 1. Let conditions $1^0-4^0$ be fulfilled and inepualihes (4)
fold.
Then the following estimete is valid

\begin{equation}
Re\left(\int\limits_{\tau}^{t} p^{-1}(\eta )\,d\eta \right) \leq
(t-\tau )\left| H(t,\tau )\right| \cos \arg H(t,\tau )\leq (t-\tau
)\left| H(t,\tau )\right| \sin \delta ,
\end{equation}

here

$$
\left| \arg H(t,\tau )\right| \leq \frac \pi 2-\delta
,\,\,\,0<\delta <\frac \pi 2,\,\,\,t\geq \tau ,\,\,\,H(t,\tau
)=\frac{1}{t-\tau} \int\limits_{\tau}^{t}P^{-1}(\eta)\,d\eta
$$

Lemma 2. Let conditions $1^0-4^0$ be fulfilled for some $\tau_0
\in \left[0,\infty \right)$, $Re\,p(\tau_0)>0$ then the estimate
\begin{equation}
Re \left(\int\limits_{\tau_0}^{t} p^{-1}(\eta)\,d\eta \right) \leq
(t-\tau _0)\left| H(t,\tau _0)\right| \sin \delta
\end{equation}

where

$$
0<\delta <\frac \pi 2,\,\,\,\left| H(t,\tau _0)\right| >0
$$

Lemma 3. Let conditions $1^0-4^0$ be fulfilled . Then the estimate

\begin{equation}
Re\left(\int\limits_{0}^{t}p^{-1}(\eta )\,d\eta \right) \leq
t\left| H_1(t)\right| \sin \delta
\end{equation}
where
$$
0<\delta <\frac \pi 2,\,\,\,\,H_1(t)=\frac{1}{t}
\int\limits_{0}^{t}p^{-1}(\eta)\,d\eta
$$
is valid.

Theorem. Let conditions $5^0, 6^0, 7^0$ and the conditions of lemmas 1, 2,
3 be fulfilled. Then problem (1), (2) has a classical solution
belonging to the space $C^{1,2}(t>0,\,\,x\in (-\infty ,\infty
))\frown C(t\geq 0,\,\,\,\,x\in (-\infty ,\infty ))$ and this
solution is represented by formula (3).

\end{document}